# Semantic system for searching of employees

MARIYA EVTIMOVA-GARDAIR, TASHO TASHEV
Faculty of Computers systems and control, English Language Faculty of Engineering
Technical University- Sofia
bul.Kliment Ohridski,8
BULGARIA
mevtimova@tu-sofia.bg, t_tashev@tu-sofia.bg

*Abstract:* Many people have stress to leave their job and start a new one because of the new environment and not enough knowledge about the culture and structure about the new organization they are going to work in. New employees in company normally need to integrate in their working place environment quicker to start doing their job. That makes them ask a lot of questions to their colleagues and sometimes their colleagues are too busy to answer those questions. In the literature is defined that this problem could be solved when new employees use digital system for information as the proposed system for searching of information. Furthermore, the quality of the returned results from the searching system is defined as a standard for the efficiency of the searching systems. Because of this, it is proposed a semantic system for searching information of employees in a company that will help to better orient new employees in a company, to know the position and the function of each employee in the company.

## 1  Introduction

Many organizations did not propose a service for the new employees to their workplaces and their jobs. Furthermore, paper working instructions and manuals are not sufficient enough for the new employees in large organizations [1], [2].

New comers in the company often ask a lot of questions the other more experienced employees in the organization and this could be quiet boring about them, so sometimes new comers in the organization are often left to survive. As a result the new employee is often confused that have result to reducing their productivity and they mostly leave the organization during the first year. Because of this, performing the effective plan continues to be crucial to increase productivity and retention of the employees. Mostly programs for new comers include values, history and employees' hierarchy. Carefully developed orientation program get new comers in the organization to implement in the company faster and the new employees are motivated to stay in the company [3].

A proper orientation program for the new employees can reduce cost, can perform positive and motivated worker and reduce the need of Supervisor [4].

Most of the people decide to change their job that is because they do not feel welcome in the organization they are working. The statement of continuous learning gives the opportunity the employees to ask questions about the problem they have or they have to resolve.

Because of this problem, in the paper is proposed a system suitable for e-orientation of the new employees in the organization. The system integrate modern technologies for searching of information as agent technologies, semantic web and big data, so that the system to return qualified results to the user.

## 2  Challenges in searching systems

The problem with increasing the quality of the returned search results is demanding nowadays, because the volume of the information increases significantly every day. Consequently that makes difficult to process the request and return the correct results. Also vague and uncertain query request also influence the quality of the returned results to the user. A lot of new reports that describe systems for searching of information include both agent technologies and ontologies for improving search engine performance. In the last few years in the literature sources are described a number of applications that use agent technologies for semantic searching of information.

## 3 Description of the semantic system for searching of employees
### 3.1 Conceptual model of the system

The aim is to create semantic search system for employees suitable for big data. The proposed conceptual model help the employees to find semantically the requested information Fig.1.

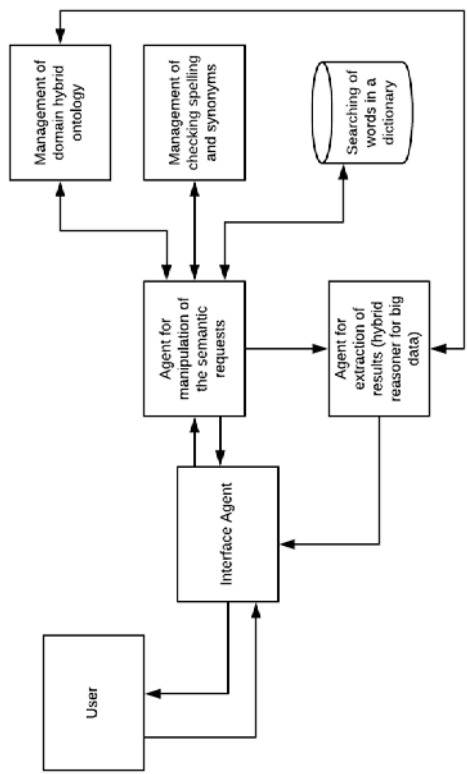

Fig.1- Conceptual model of semantic search system of employees

This conceptual model represent semantic searching of employees that are responsible for certain job from the new comers in the organization. This platform help new comers to find the employee in the organization responsible for a certain work and present to the user of the system his or her name, phone e-mail and the title of the position. This system will give to the new employees' quick information to finish their tasks without bothering and disturbing the other employees that have more experience within the organization. That platform will help the new comers to be able more easily to enter in the organization and to start doing their job efficiently. In the nowadays, new methodologies to improve searching of the information mostly use semantics. Semantics search system scientifically proves more qualified returned results to the user. Because semantic provide searching using the meaning of the words it is used ontology [5].

Moreover in recent years there has been growing interest in investigating multi- agent systems for searching of information [6]. Nowadays multi-agent system search is mainly focused on coordinating agent behavior and distributing and unification of decisions that find place in systems for searching of information [7],[8] especially when searching information in big volume of data.

The agent has an effective communication and abstraction mechanism, because of this it is recommended to use agents [9]. The proposed conceptual model consists from three agents: Interface agent, that is responsible for all the interactions with the user, Agent for manipulation of the semantic request, that manipulate the user requests and Agent for extraction of the results, that use hybrid reasoner suitable for big data [10] when performing reasoning from the hybrid ontology.

The functionality of the system is shown in Fig.2.

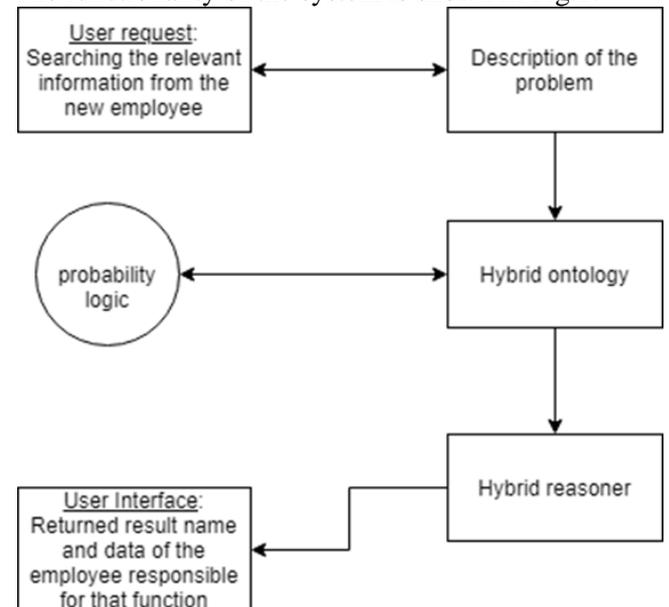

Fig 2- Description of the semantic search system for employees

It is used predefined domain hybrid ontology for a company that sells products, in order to be able to understand semantically the context of the questions request from the user Fig.3,Fig.4.

In Fig.4 to define the object properties of the ontology suitable for big data is added hasPeer and hasFactor [11]. This information is added using the plug-in Jess in Protégé. The algorithm of reasoning is described in [12].

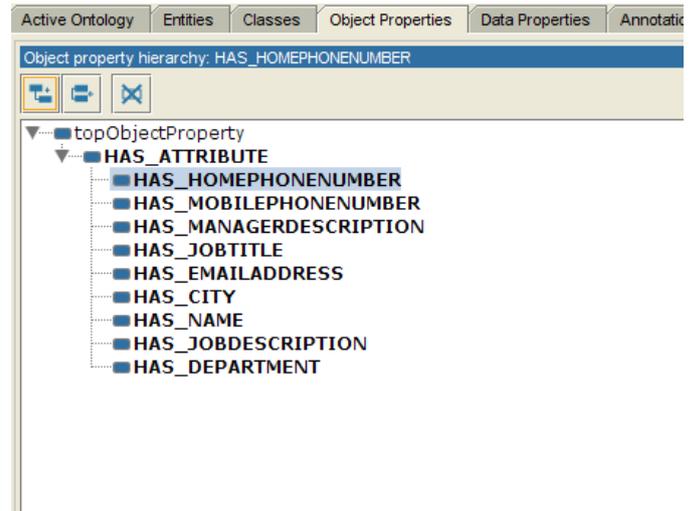

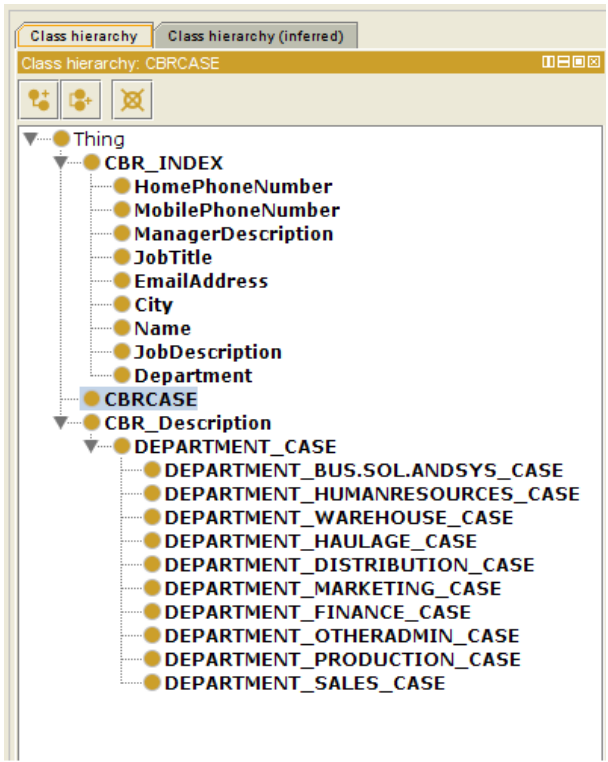

Fig.3- Description of the classes in the ontology

When extending the object properties of the ontology to work with big data[13]:

hasFactor- represent factor of imperfection for each rule or case

hasPeer- represent number of peers that are related with each rule or case or with other key value

As a key value that could be the different department of the company.

Fig.4-Case based description of the object properties in the ontology

Ontology can be defined as a formal presentation of knowledge, "a set of concepts within a domain, using a common vocabulary to show the types, properties and relationships of these concepts". The user is entering free query as it is using user interface. After that the query is manipulated and after reasoning the results is returned to the user.

## 3.2 The ontology integrated into the application

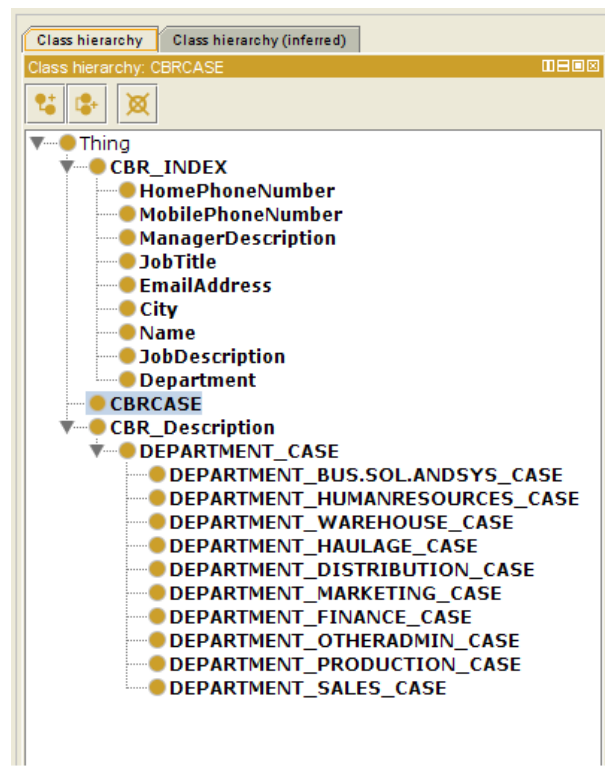

Fig.3- Description of the classes into the ontology

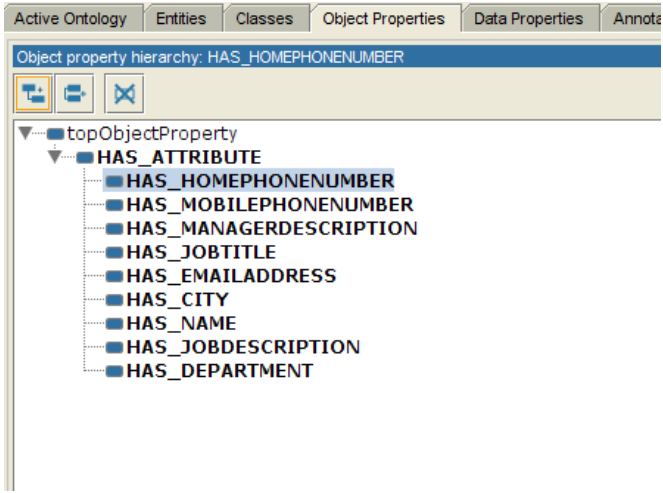

Fig.4- Case based description of the object properties into the ontology

## 4 Evaluation and performance

To develop the software it is used the java library jcolibri, Jess and Protege.

Basically to increase the quality, when searching with ontology is to maximize Precision, Recall and F-measurement. To create the performance it is gathered 216 different queries from employees of the company. Then the queries are categorized concerning the department of the request.

The results for Precision in each department is shown in Fig.5.

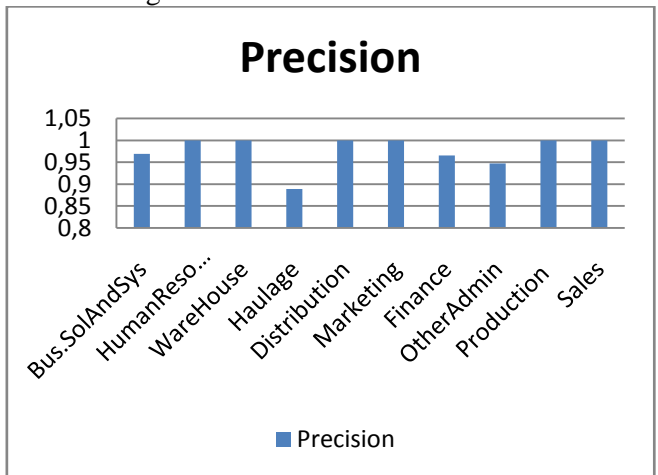

Fig.5- System measurement with Precision

The results for measurement of the system Recall is shown in Fig.6.

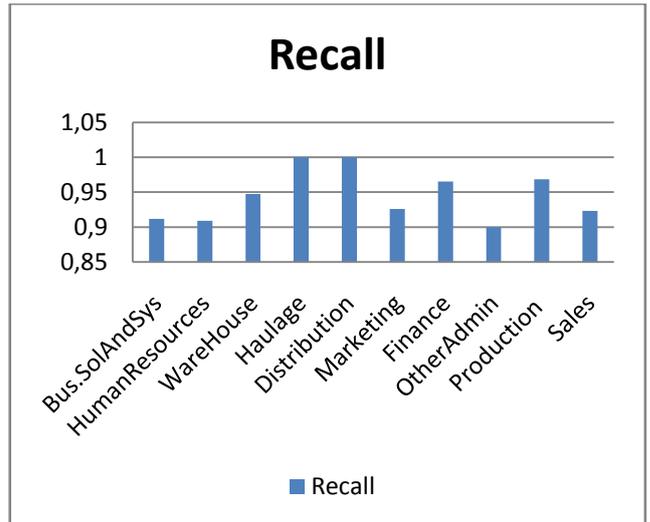

Fig.6- System measurement with recall

The result for measurement of the system with F-measure is shown in Fig.7.

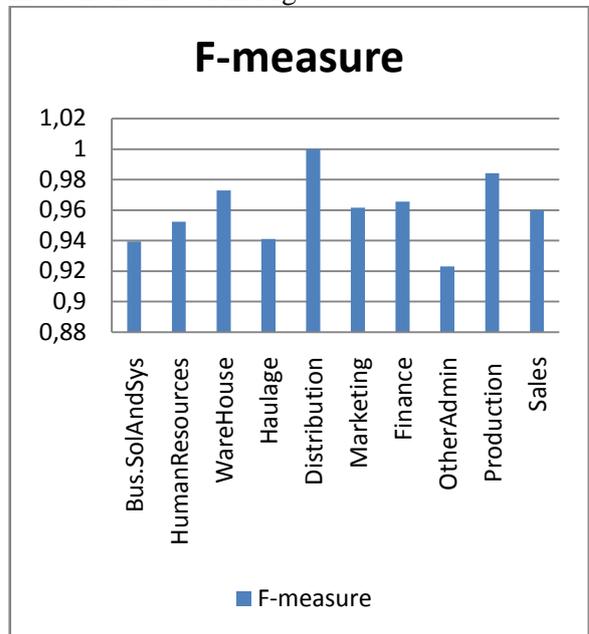

Fig.7.- System measurement with F-measure

As a result the system have 98% Precision, 94%Recall and 96% F-measure.

## 5 Conclusion

A model of semantic searching system of employees is proposed. It is described an ontology for employees that is case- based and is suitable for big data. It is described a new model of ontology that is case- based and is appropriate for big data. The system that is described is suitable for company for products but could be adapted for another company. The software is tested in a company for bottling drinks and gives reasonable qualified results with precision 98%, that show the quality of the returned results from the system. The results from the system are relatively high than the other returned results from the systems described in the literature. In the future the system have to be measured with data from other companies.